\documentclass[12pt]{article}
\usepackage{docident}
\usepackage{latexsym}
\title{Quantum-type Coherence as a Combination of Symmetry and Semantics.}
\author{Yuri F. Orlov}

\begin{document}

\maketitle
\docident{\hspace{\fill}\makebox[0pt][r]{\sf CLNS 97/1476}}

\center{Floyd R. Newman Laboratory of Nuclear Studies\\
Cornell University, Ithaca, New York 14853 USA}

\abstract{It is shown that quantum-type coherence, leading to
indeterminism and interference of probabilities, may in principle
exist in the absence of the Planck constant and a Hamiltonian.  Such
coherence is a combined effect of a symmetry (not necessary physical)
and semantics.  The crucial condition is that symmetries should apply
to logical statements about observables.  A theoretical example of a
non-quantum system with quantum-type properties is analysed.}

\bigskip

\medskip

\medskip

\indent{Coherence, a cornerstone of quantum mechanics, is considered
to be a result of the quantization of action.  However, we will show
that ``quantum-type coherence,'' as we will call it, does not depend
on the existence of the Planck constant (although its concrete
manifestations do).  In our two examples, $E_{1}$ and $E_{2}$,
analysed below, such coherence appears in the presence of:(I) An
ordered set of mutually exclusive objects, with numerical values
$\xi,\xi\in$A, where A is an affine space (a space with no fixed
origin); $\xi$ is a particle coordinate in $E_{1}$, and an
interpretation of a given situation in $E_{2}$. $\linebreak$
(II) Another ordered set of mutually exclusive objects, with numerical 
values $\chi$ defined on set (I) {\it as a whole}; $\chi$ is a value of
the particle momentum in $E_{1}$, and in $E_{2}$ the ordinal
number of a logical statement in a set of mutually exclusive
statements describing the situation.  Conditions(I) and (II)
imply the existence of a symmetry. It is shown that when symmetries
apply to logical statements about objects instead of objects ``per se,'' 
the following semantic problems arise: in $E_{1}$, the problem of
expressing the truth values of logical statements about objects in the
second set, in terms of the truth values of logical statements about
objects in the first set; and in $E_{2}$, the problem of expressing
the truth values of statements in one interpretation when the truth
values of the same statements in another interpretation are given.
Inexpressibility is therefore a combined effect of symmetry and
semantics---both irrelevant to $\hbar$. This effect, fundamental for
quantum-type coherence, leads to indeterminism and interference of
probabilities.}

\medskip

\indent{We will first delineate the border between the
symmetry-semantic part (without $\hbar$) and the quantum part (with
$\hbar$) of the quantum mechanical formalism, using an example of a
single, zero-spin particle.  Then we will analyse a non-quantum system
having all typical features of quantum-type coherence.  Its analysis
provides a theoretical basis for searching for systems, neither
classical nor quantum, in which quantum-type interference can be
observed.  Observation of such systems, interesting in itself, may
indirectly clarify our understanding of the structure of quantum
mechanics and the origins of quantization.}

\medskip

\indent{$E_{1}$.{\it A quantum example}.  We will construct the
quantum mechanics formalism of a single, zero-spin particle,
introducing assumptions step-by-step so as to make clear exactly at
which point in our construction $\hbar$ is needed.  We will tag our
assumptions with Greek letters.}

\medskip

\indent{Our fundamental assumption about what makes mechanics
``quantum`` is that $(\alpha)$ the statement $\Lambda_{p_0}:``p=p_{0}``$
about a particle momentum $p$ (a translational invariant in the
coordinate $q$-space) should itself be an invariant of the
translational symmetry in the same space.  Such a requirement makes
sense only if $\Lambda_{p_{0}}$, a logical statement, is
simultaneously a function of coordinates, such that it does not depend
on transformations $q\to q+\delta q$.  This is possible only if
$\Lambda_{p_{0}}$, as a function of coordinates, either does not
depend on $q$ at all, or depends only on differences between
coordinates.  We suppose the latter,
$(\beta)\Lambda_{p_{0}}(q,q')=\Lambda_{p_{0}}(q'-q)$.  In our next
step, $\Lambda_{p_{0}}(q,q')$ is assumed to be a matrix in $q$-space;
and since $\Lambda_{p_{0}}$ is a logical statement, we can use logic
to calculate it.  Consider the logical equivalence,
$\Lambda_{p_{0}}\wedge\Lambda_{p_{0}}\sim\Lambda_{p_{0}}$.  The
natural assumption is that $(\gamma)$: the logical conjunction in this
equivalence should be represented by the matrix product, and the
logical equivalence by the equation:

\begin{equation}
\int\Lambda_{p_{0}}(q'-s)\Lambda_{p_{0}}(s-q)ds=\Lambda_{p_{0}}(q'-q).
\end{equation}
The solution of this equation is $\Lambda_{p_{0}}=L^{-1}exp2\pi
i(q'-q)\lambda^{-1}(p_{0})$, where the wave length $\lambda(p_{0})$ is
an unknown function of $p_{0}$, and $L\equiv\int dq$.  The
eigenvectors of this matrix (up to normalization constants) are
$\psi_{p}(q)=exp2\pi iq/\lambda(p), \lambda(p_{1})\neq\lambda(p_{2})$
for $p_{1}\neq p_{2}$.  Thus, we already have---without $\hbar$---the
correct wave functions and density matrices, though the dependence of
$\lambda$ on $p$ remains unknown.  To be consistent, we now assume
that a statement ``$q=q_{0}$'' about particle coordinate $q$ should
also be represented in $q$-space by a matrix.  Obviously this matrix
must be
$(\delta):\Lambda_{q_{0}}(q,q')=K\delta(q'-q_{0})\delta(q-q_{0})$,
with $K$ constant.  The eigenvectors of $\Lambda_{q_{0}}$ are
$\psi_{q_{i}}(q)=\delta(q-q_{i}).$}

\medskip

Matrices $\Lambda_{q_{0}}(q,q^{\prime})$ and $\Lambda_{p_{0}}(q,q^{\prime})$
do not commute; their commutator is not proportional to $\hbar$.
While $\Lambda_{q_{0}}$ is a statement about the exact location of the
particle, statement $\Lambda_{p_{0}}$ expresses {\it by its own symmetric 
structure} the uncertainty of that location, and is defined by the
momentum value, $p_{0}$, which, according to the meaning of
translational invariance, relates to $q$-space as a whole.  If, now,
$\Lambda_{p_{0}}$ is true, i.e., is the correct description of the
state of the particle, and the question is whether $\Lambda_{q_{0}}$
is true, the answer can be at best probabilistic, since the truth of
$\Lambda_{p_{0}}$ is inexpressible in terms of the truth of any
$\Lambda_{q_{i}}$.  Here we have the probability that the conjunction
$\Lambda_{p_{0}}\wedge\Lambda_{q_{0}}$ is true.  Since, according to
$(\gamma)$, conjunctions are represented by matrix products, and since
the probability should be a translational invariant and be independent
of the order of matrices, the only correct formula is:

\begin{equation}
w(p_{0}\vert q_{0})=\int\Lambda_{p_{0}}(q,q')\Lambda_{q_{0}}(q',q)dqdq'=K/L.
\end{equation}
Only after this step in constructing the quantum formalism need we
introduce $\hbar$.  The physical part of the quantum formalism is
then defined by the introduction of Hamiltonian, canonical
transformations, etc.

\medskip

{\it E$_{2}$. A non-quantum example.} Consider a situation open to
interpretation and describable by different logical connections among
{\it n} independent logical statements, $\lambda_{i},
i=1,2,...,n$.  According to the classical logic of propositions,
every such description can be represented by a disjunction of mutually
exclusive conjunctions.  There are $N=2^{n}$ such conjunctions, which
can be enumerated as $\Lambda_{k}, k=1,2...,N; N\geq 2$.  One, and only
one of them, can be true.  By definition, a {\it certain} (i.e., not
uncertain) interpretation $I^{(s)},I^{(s)}=I^{(s)}(k),$ is the following
function of the integer $k, k=1,2,...N$: if some $\Lambda_{l}$ is
defined as a true statement-conjunction, then
$I^{(s)}(k)=\delta_{kl}$.  Inversely, if $I^{(s)}(k)=\delta_{kl},$
then in this interpretation $\Lambda_{l}$ is true.  Two
interpretations are identical if the corresponding functions are
identical.  There are only {\it N} non-identical certain
interpretations, each defining which {\it one} of the {\it N}
conjunctions is true.  {\it N} non-identical certain interpretations
can be transformed into each other by permutations, described by the
finite table.  That table can be considered as an algorithm defining
the truth values of the conjunctions in all certain interpretations,
when the truth values of the conjunctions in one of them are defined.

\medskip

Now we will extend our concept of interpretation beyond classical
logic by introducing the following two conditions: (a)(symmetry) There
is no correct interpretation a priori: any statement-conjunction may
be considered true.  Such a choice defines a {\it correct}
interpretation.  Once an interpretation is considered correct, {\it N}
different certain interpretations can be generated by permutations of
truth values.  By definition, these certain interpretations define the
meaning of the truth values of {\it N} conjunctions. (b) If there are
two interpretations, $I^{(s)}(k)$ and $I^{(s')}(k)$, then the difference
between them is measured by a real number, $\theta$; all values of
$\theta$ inside an interval $\theta_{min}\leq\theta\leq\theta_{max}$
are permitted; and if $I^{(s)}(k)$ is considered correct, then $\theta
=s'-s$.

\medskip

More formally, all interpretations are now points in an affine space
$A^{1},s\in A^{1}$, and $\vec\theta$ is a vector in $R^{1}$ vector
space of real numbers, $\theta \in R^{1}$.  The group $R^{1}$ acts on
$A^{1}$ as the continuous group of parallel displacements.  Points of
our affine space are {\it functions} defined in their own discrete
spaces, each containing {\it N} points, $N\geq 2$.  We will show that
such a system possesses the properties of quantum-type coherence:
indeterminism, interference of probabilities, and the possibility of
introducing wave functions, though none of our assumptions depends on
$\hbar$.  The reason, qualitatively, is that now we have a continuum
of interpretations; but when we define a meaning of truth values of
{\it N} conjunctions that is equal for all interpretations, only {\it
N} interpretations (which, according to (a), can be chosen
arbitrarily) can be certain.  In all other interpretations, truth
values of conjunctions are not certain.

\medskip

{\it Theorem 1. On the existence of inexpressibility.}

\medskip

{\it If conditions (a) and (b) are met, then either all
interpretations are identical or there does not exist any algorithm,
defined for $\theta$ in the interval $[\theta_{min},\theta_{max}]$, to
calculate the truth values of statements in an interpretation
$I^{(s')}(k)$ when some other interpretation, $I^{(s)}(k)$, is
considered correct}

\medskip

{\it Proof.} Let statements $\Lambda_{l}, l=1,2,...,N,$ in some
interpretation,$I^{(0)}(l),$ that is considered correct, possess given
truth values, and let not all interpretations be identical.  Then
there exists some $\theta$ that defines an interpretation,
$I^{(\theta)}(k),$ different from $I^{(0)}(k)$.  This means that at
least one of the statements in $I^{(\theta)}$, let it be
$\Lambda_{i}$, does not possess the same truth value as $\Lambda_{i}$
in $I^{(0)}$.  Assume that there exists an algorithm mapping the truth
values of statements in $I^{(0)}$ onto the truth values of statements
in $I^{(\theta)}$. Since two different distributions of truth values
among {\it N} statements are permutations of each other, any assumed
algorithm should define the operation of a permutation, which should
depend on $\theta$.  Consider the parameter $\delta\theta=\theta/N$!
connecting any two interpretations with $s'-s=\delta\theta$.
According to our assumption, this parameter defines some permutation
of truth values.  Consider {\it N!} such consecutive permutations,
beginning from the given distribution of truth values in the initial
interpretation, $I^{(0)}$. On one hand, {\it N!} identical
permutations give us the same final distribution of truth values as
the initial one.  But, on the other hand, since $\delta\theta\cdot
N!=\theta$, we will arrive at interpretations $I^{(\theta)}$, which is
not identical to the initial interpretation.  The contradiction means
that our assumption about the existence of a mapping algorithm
dependent on $\theta$ was wrong.  It also means that not all
interpretations can be certain.  The truth values of {\it N}
conjunctions in uncertain interpretations cannot be expressed in terms
of the truth values of these conjunctions in certain
interpretations. $\Box$

\medskip

Thus, under the conditions satisfying Theorem 1, the problem of
expressing the truth values of statements in an arbitrary
interpretation, when the truth values of statements in another,
certain interpretation are given, is insoluble.  Therefore, in the
general case, if a question arises whether a particular statement in
an interpretation is true when the truth values of statements in
another interpretation are given, the answer is unpredictable.  If the
concept of probability applies to such a system, then the probability
of a certain answer can depend only on the parameter defining the
difference between interpretations, $\theta$.  This leads to

{\it Theorem 2. Under the conditions of Theorem 1, the probabilities
p$(\theta)$ of complex events do not obey classical rules, and have
interference terms.}

\medskip

{\it Proof.}It is sufficient to prove the theorem for the simplest
case {\it N}=2, in which there are only two mutually exclusive
statements in every interpretation, $\Lambda$ and
$\overline{\Lambda}$; the latter is the negation of the first.
Consider three pairs of interpretations, ${\left( I^{(0)},I^{(\theta)}\right)
\left(,I^{(\theta)},I^{(\theta + \vartheta)}\right)}$, and ${\left(
I^{(0)},I^{(\theta + \vartheta)}\right)}$. We will label s the statement,
either $\Lambda$ or $\bar{\Lambda}$, whose truth values are
interpreted in the (maybe uncertain) interpretation $I^{(s)}(k),
k=1,2;\;\Lambda_{1}\equiv\Lambda, \Lambda_{2}\equiv\overline{\Lambda}$.
Probabilities of the answer ``yes'' to the questions ``Is
$\Lambda^{(r)}$ (or $\bar{\Lambda}^{(r)})$ true, if $\Lambda^{(s)}$ is
true?'' will be denoted as $p(s,r)\equiv p(r-s),$ and
$p(s,\bar{r})=1-p(s,r)$; and the probabilities of the answer ``yes''
to the same questions but with $\bar{\Lambda}^{(s)}$ instead of
$\Lambda^{(s)}$ will be denoted as $p(\bar{s},r)=1 - p(s,r)$, and
$p(\bar{s},\bar{r})=p(s,r)$.  The equalities follow from the fact that
in any interpretation, certain or not, one of the statements is true
and the others are false, because the disjunction
\begin{equation}
\Lambda_{1}\vee\Lambda_{2}\vee ...\vee\Lambda_{N}\equiv T
\end{equation}
is an invariant (a tautology) independent of the choice of interpretation.

\medskip

Given questions corresponding to the aforementioned three pairs of
interpretations, such that either $\Lambda^{(s)}$ or
$\bar{\Lambda}^{(s)}$ is true in interpretation $I^{(s)}$ of every
pair, the classical probability formula for the answers should be:
$p(0,\theta + \vartheta)=p(0,\theta)p(\theta,\theta +
\vartheta)+p(0,\bar{\theta})p(\bar{\theta},\theta + \vartheta),$
which can be rewritten as
\begin{equation}
p(\theta + \vartheta)=p(\theta)p(\vartheta)+(1 - p(\theta))(1 -
p(\vartheta)),\;\; classical.
\end{equation}
But this equation is violated, for example, when $p(\theta +
\vartheta)$=0, i.e., when $(\theta + \vartheta)$ is such that in
interpretation $I^{(\theta + \vartheta)}$ ``yes'' (``no'') means the same
as the ``no'' (``yes'') of interpretation $I^{(0)}$.  Choosing $\theta =
\vartheta$, then, gives us $0 = p^{2} + (1 - p)^{2}$, and this is
impossible.  Therefore, there must be an additional term in (4).
Moreover, in our simple case we can calculate this term under the
assumption
\begin{equation}
p(0,\theta)=p(\theta,0).
\end{equation}
Choosing $\vartheta =-\theta$ gives us another classical equation:
$1=p^{2}+(1-p)^{2}$, which can be rewritten as
\begin{equation}
1=cos^{4}f(\theta)+sin^{4}f(\theta),\;\; classical,
\end{equation}
where the function $f(\theta)$ needs to be found.  In this case, the
needed additional term should be
$2sin^{2}f(\theta)cos^{2}f(\theta)$. From this we can find
\begin{equation}
f(\theta)=a\theta
\end{equation}
where $a$ is an arbitrary real number.  Indeed, from (7) it follows that
\begin{equation}
p(\theta +\vartheta)\equiv cos^{2}a(\theta +
\vartheta)=p(\theta)p(\vartheta)+(1-p(\theta))(1-p(\vartheta))+ 
interference\;\;term,
\end{equation}
\begin{equation}
interference\;\;term=-2sin a\theta\;\; sin a\vartheta\;\; cos a\theta\;\; cos a
\vartheta.
\end{equation}
This formula gives correct results in cases $\vartheta =-\theta,$ and
$a\vartheta = a\theta =\pi/4,$ while deviations from (7) do
not.$\Box$

\medskip

When $a = 1/2$, formulae (8),(9) coincide with the quantum formulae
for a spin 1/2 placed on a plane, with the rotational symmetry around
the axis perpendicular to this plane; $\theta$ is an angle between two
axes, $z$ and $z'$, placed on this plane; $\Lambda^{(z)}$ is the
statement $``s_{z}=1/2``$; and $\bar{\Lambda}^{(z)}$ the statement
$``s_{z}=-1/2.``$ From this analogy it is clear that we can introduce
formal wave functions as superpositions of certain interpretations, as
defined above, thus introducing a Hilbert space. (We will not do this
here.)  There is an essential difference, however, between our
non-physical system and a quantum mechanical one.  In quantum
mechanics, the discreteness of spin $z$-projections is a direct result
of $SU$2 symmetry in a three-dimensional space; such discreteness would
not exist in a system with a rotational symmetry only on a plane.  In
our logical system, the discreteness---which is simply the discreteness
of logical truth values---exists independently of symmetries.

\end{document}